\documentclass[prl, twocolumn, superscriptaddress, footinbib, showkeys]{revtex4-2}

\usepackage[english]{babel}
\usepackage{lipsum}

\usepackage{natbib}
\usepackage{url}

\usepackage[pdftex]{graphicx}
\usepackage{subfigure}
\usepackage{color}
\usepackage[toc,page]{appendix}
\usepackage{amsmath}
\usepackage{amssymb}
\usepackage{mathtools}
\usepackage{braket}
\usepackage{stmaryrd}
\usepackage{xfrac}

\usepackage{float}

\usepackage{verbatim} 

\usepackage[sort&compress]{natbib}
\usepackage[unicode]{hyperref}
\hypersetup{
    colorlinks=true,
    linkcolor=blue,
    citecolor=magenta,
}

\begin{document}

\title{Impact of impurities on the topological boundaries and edge state localization in a staggered chain of atoms: SSH model and its topoelectrical circuit realization}
%

\author{J.~C.~Perez-Pedraza}
\email{julio.perez@umich.mx}
\affiliation{Instituto de F{\'i}sica y Matem{\'a}ticas, Universidad Michoacana de San Nicolás de Hidalgo, Edificio C-3, Ciudad Universitaria, Francisco J. M{\'u}jica S/N Col. Fel{\'i}citas del R{\'i}o, 58040 Morelia, Michoac{\'a}n, M{\'e}xico.}

\author{J.~E.~Barrios-Vargas}
\email{j.e.barrios@gmail.com}
\affiliation{Departamento de Física y Química Teórica, Facultad de Química, Universidad Nacional Autónoma de México, Ciudad de México 04510, México}

\author{A. Raya}
\email{alfredo.raya@umich.mx}
\affiliation{Instituto de F{\'i}sica y Matem{\'a}ticas, Universidad Michoacana de San Nicolás de Hidalgo, Edificio C-3, Ciudad Universitaria, Francisco J. M{\'u}jica S/N Col. Fel{\'i}citas del R{\'i}o, 58040 Morelia, Michoac{\'a}n, M{\'e}xico.}
\affiliation{
 Centro de Ciencias Exactas - Universidad del Bio-Bio. Avda. Andr\'es Bello 720, Casilla 447, Chillán, Chile.
}

\date{\today}

\begin{abstract}
We study the Su-Schrieffer-Hegger model, perhaps the simplest realization of a topological insulator, in the presence of an embedded impurity superlattice. We consider the impact of the said impurity by changing the hopping amplitudes between them and their nearest neighbors in the topological boundaries and the edge state localization in the chain of atoms. Within a tight-binding approach and through a topolectrical circuit simulation, we consider three different impurity-hopping amplitudes. We found a relaxation of the condition between hopping parameters for the topologically trivial and non-trivial phase boundary and a more profound edge state localization given by the impurity position within the supercell. 
\end{abstract}


\maketitle

\section{Introduction} \refstepcounter{section}
The emergence of topological matter has challenged the point of view of the classification of quantum materials in terms of their symmetry properties as described by unitary transformations and opened the possibility of a wider classification. In that direction, number of studies have been carried out to understand the mechanisms responsible for topological properties \cite{von1986quantized, klitzing1980new, stormer1999fractional, kane2005z,Batra2020,2016LNP...919.....A, 10.1088/978-1-68174-517-6}. Time-reversal, particle-hole, and chiral symmetries, namely, unitary or antiunitary transformations that commute or anticommute with the Hamiltonian, are the building blocks of the so-called periodic table of topological insulators~\cite{periodic}. 
The \textit{Su-Schrieffer-Heeger} model (SSH)~\cite{ssh,PhysRevB.22.2099} is one of the simplest models which represents a topological insulator (see, for instance, Refs.~\cite{RevModPhys.82.3045,RevModPhys.83.1057,shankar2018topological, Batra2020, wang2013topological, heeger1988solitons, yu1988solitons, ryu2010topological} and references therein). It describes the behavior of spinless electrons hopping through a one-dimensional lattice build up by two interspersed sublattices of atoms with alternating nearest neighbors (NNs) {\it{hopping}} amplitudes. As first introduced, this model is useful for studying one-dimensional molecules such as polyacetylene $(CH)_x$. Its topological features, nevertheless, have boosted the interest in studying it and its extensions in a variety of situations, such as modulations of the hoppings and on-site energies (driven SSH model \cite{peng2016experimental, gomez2013floquet, dal2015floquet}), long-range interactions \cite{an2018engineering, perez2019interplay, li2018topological}, two coupled SSH chains \cite{li2017topological, junemann2017exploring, sun2017quantum}, dimensional extended models \cite{li2022dirac, xie2019topological, li2014topological, agrawal2023cataloging}, and other modifications \cite{li2022topological, sivan2022topology}. Depending upon the relative strength of the hopping parameters, the model exhibits a topologically trivial or nontrivial phase, which is distinguished by the
emergence of a zero mode in the spectrum. The topologically invariant quantity turns out to be the Zak phase, which is either zero or one in the trivial and nontrivial phases, respectively~\cite{zak1989berry}. These features make the SSH a favorite model to predict a nontrivial topological structure of some systems from the symmetries of the underlying Hamiltonian and the corresponding equations of motion. 

The topological features of the SSH model have also been found in mechanical \cite{betancurocampo2023twofold}, photonic \cite{wang2018experimental, chen2018observation}, acoustic \cite{xiao2015geometric, peng2016experimental} and other systems \cite{stuhl2015visualizing, meier2016observation, meier2018observation, cai2019experimental, goldman2016topological, jotzu2014experimental}. The basic idea is that these {\em metamaterials} can be described by a two-band Hamiltonian, which yields similar equations of motion in the tight-binding regime as the model in question. In this regard, in the emergent field of topoelectric circuits~\cite{PhysRevB.100.165419,doi:10.1142/S2529732520970020,RevModPhys.93.025005,Lee2018,ZHAO2018289,PhysRevResearch.3.023056} one can map the current flow in electric circuits with a network of passive elements like capacitors and inductors in similar form to a tight-binding Hamiltonian where capacitances C and inductances L serve to define the hopping parameters. Hence, crystal systems like the SSH model, graphene and others with remarkable properties have found realization within these topoelectric circuits \cite{xun2020topological, Lee2018, ningyuan2015time, zhu2018simulating}. For the case of the SSH model, an alternating network of capacitors and inductors play the role of a unit cell in a crystal whereas products of capacitances and inductances define the hopping parameters. 
The impedance of the system as generated by a resistor probe and a sinusoidal signal with varying frequency can be used to measure the response of the system.  In the topological phase, such an impedance is found to diverge at a certain frequency in the topological phase of the system. This resonant regime can be expressed entirely in terms of the hopping parameters of the crystal. Moreover, localization of states can be visualized in terms of edge states formation in the topologically non-trivial phase. This opens the possibility of studying interesting crystallographic properties of crystals in terms of a network of LC circuits~\cite{PhysRevB.100.165419,doi:10.1142/S2529732520970020,RevModPhys.93.025005,Lee2018,ZHAO2018289,PhysRevResearch.3.023056,xun2020topological, ningyuan2015time, zhu2018simulating}. 
\begin{figure*}[ht]
   \includegraphics[width=0.85\textwidth]{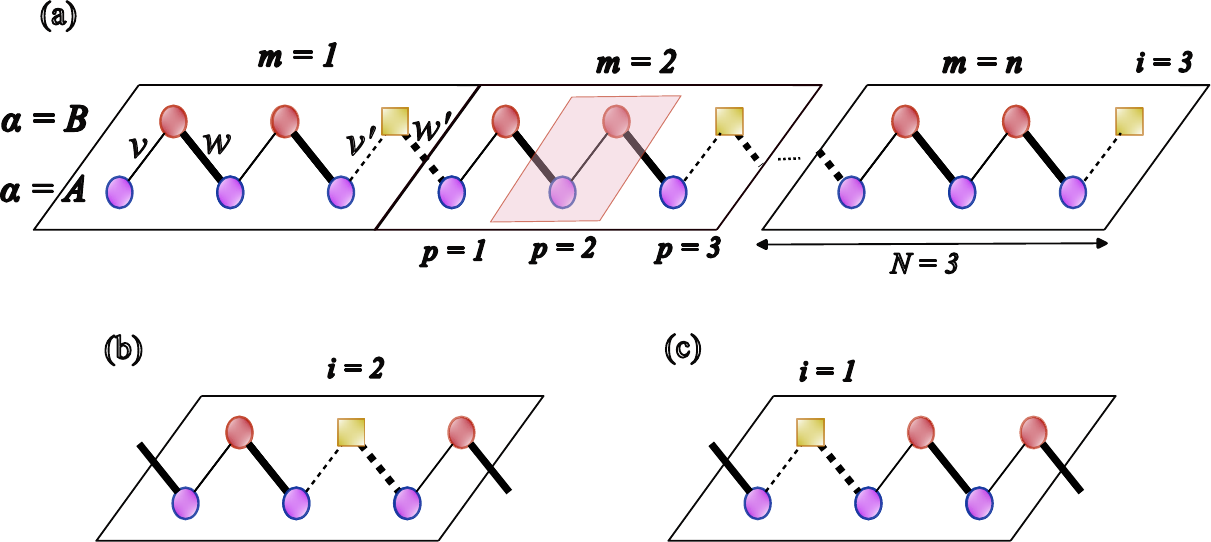}
   \caption{(a) Scheme of the $N,i$-super-SSH model for $N=3$, $i=3$. The shaded area denotes the SSH-unit cell, the supercell number is given by the parameter $m$, the supercell position by $p$, the species parameter is $\alpha$ and the impurity position is denoted by $i$. (b) and (c): Supercells of the case $N=3$, $i=2$(b) and $i=1$(c), respectively. In all cases the green rectangles indicate the impurity atoms, the solid thin and thick lines correspond to the hopping parameters $v$ and $w$, respectively, while the dashed thin and thick lines denote the  $v'$ and $w'$ impurity hopping parameters, respectively.}
    \label{fig:mod_super}
\end{figure*}

Of special interest is the role of impurities in the system from both the crystallographic and the electronic point of view. In this article, we explore the role of an impurity into the system in different setups by defining the $N,i$-super-SSH model. First, we consider an impurity in a super-cell that interacts with the network and observe its impact on the position of the resonance in the impedance. Then, we study the impact of its position within the sites of the super-cell. We explore the phase diagram in the parameter space of the hoppings of the chain and the emergence of edge states in the topological phase of the system. We analyze the dynamics both from the tight-binding perspective and then from an electric circuit simulation.
The rest of this article is organized as follows. In the next section, we present the model under consideration. In section~\ref{sec:TB}, we analyze the system in the Tight-Binding (TB) approach through the Python package ``Pythtb'' (Python Tight-Binding)~\cite{pythTB}. In Section~\ref{sec:circuit}, we perform a numerical simulation of a topoelectrical circuit equivalent to our model. We conclude in Section~\ref{sec:conclusions} and present some details of the framework of topological circuits in an Appendix.

\section{SSH model with embedded impurity superlattices}\refstepcounter{section}\label{sec:model}

Let us begin our discussion by considering a one-dimensional chain of alternating atoms of species $A$ and $B$, from which we define the species parameter $\alpha= A,B$, and hopping parameters $v$ and $w$ for the $A-B$ and $B-A$ links, as in the standard presentation of the SSH model, in which a unit cell consists of a pair of atoms $A$ and $B$ (see shaded area in Fig.~\ref{fig:mod_super}(a)). Let us also consider a supercell constructed by concatenating $N$ SSH-unit cells. Such supercell has length $N$ and we label the sites within the supercell as $p=1,2,...,N$. Notice that each site consists of one SSH-unit cell in which one atom of species $B$ is replaced by an impurity consisting of an atom of a third species $C$ located in $p=i$, as shown in Fig.~\ref{fig:mod_super}(a) for the case $N=3$, $i=3$. The system can then be described by four principal numbers: $m = 1,2,3,...,n,...$ denoting the number of supercells in the array, $N$ denoting its length, $p$ representing the label of the site position within the supercell, $\alpha$ denoting the atomic species, and an additional label $i$ indicating the position of the impurity within the supercell. We dub this arrangement the {\em $(N,i)$-super-SSH model}. 
Figure~\ref{fig:mod_super}(a) shows cases of the $(3,3)$-super-SSH model, while Figs.~\ref{fig:mod_super}(b) and (c) show the supercells of the $(3,2)$- and $(3,1)$-super-SSH model, respectively. 

The effect of the impurities is based on the change of the hopping amplitudes between sites $A-C$ and $C-A$ to the new hoppings $v'$ and $w'$, respectively. Under this assumption, the system can still be thought as periodic, but now with a new unit supercell. In the general case, the quantities $v'$ and $w'$ can be considered as independent parameters or even functions of $v$ and $w$. Moreover, for the purposes of this study, we explore electronic and topological properties of the system considering them as scalar parameters $v'=v+\delta v$ and $w'=w+\delta w$, under different assumptions for three different cases. We consider the periodic and the finite cases varying the supercell length $N$ and the impurity position $i$ within the supercell. As a first approach, below we consider a TB description of the system.


\section{TB calculation} \refstepcounter{section}\label{sec:TB}
In this Section, we compute the electronic and topological properties of our system from a TB perspective. We write the TB Hamiltonian as
\begin{equation}\label{hamiltonian}
H_{super}= H_{SSH}^N + V_{N,i},
\end{equation}
where we define the extended SSH Hamiltonian as
\begin{widetext}
\begin{equation}
    H_{SSH}^{N} = \sum_{p=1}^{N-1} \left(v |m,p,B\rangle \langle m,p,A| + w |m,p+1,A\rangle \langle m,p,B|\right) + v |m,N,B\rangle \langle m,N,A| + w |m,1,A\rangle \langle m,N,B|,
\end{equation}
and the impurity potential
\begin{equation}
    V_{N,i} =  \begin{cases}
        \delta v |m,p,B\rightarrow C \rangle \langle m,p,A| + \delta w |m,p+1,A\rangle \langle m,p,B\rightarrow C|, \ \text{for}\ i=1,2,...,N-1, \\
        \delta v |m,p,B\rightarrow C \rangle \langle m,p,A| + \delta w |m+1,1,A\rangle \langle m,p,B\rightarrow C|, \ \text{for}\ i=N. \\
    \end{cases}\label{eq:potential}
\end{equation}
\end{widetext}
It is important to note that in Eq.~\eqref{eq:potential}, the notation $B\rightarrow C$ means the replacement of an atom $B$ by an impurity atom $C$. The exact algebraic diagonalization of the Hamiltonian in Eq.~(\ref{hamiltonian}) gives as result tremendously intricate expressions for the energy-momentum dispersion, which results very difficult to manipulate. For this reason, we employ the Pythtb package~\cite{pythTB} in order to get  the eigenvalues and eigenvectors of the problem numerically. We consider three different configurations of the impurity-hopping amplitudes: In Case I, we consider $w=1.0$ fixed and $v'=w'$, namely, equal impurity hopping to both nearest neighbours, rendering $v$ and $v'$ as the free parameters which we vary. In Case II, we explore the well-known topologically non-trivial phase of the SSH model with $w=1.0$ and $v=0.5$, and vice versa, namely, the topologically trivial phase~\cite{Batra2020,2016LNP...919.....A, 10.1088/978-1-68174-517-6}. We retain these values fixed, while varying the free parameters $v'$ and $w'$. Finally, in Case III, we consider  $w=w'=1.0$ for which the impurity does not affect the $A-C$ hopping amplitude. This permits the variation of the free parameters $(v,v')$. 

For these three cases, we first explore the topological phase of the infinite periodic system imposing periodic boundary conditions with the hopping parameters as defined in each case. In order to do so, we first divide the First Brillouin Zone (FBZ) of the system into a discrete grid of $M$ equally spaced intervals. In each point of the grid, we consider its corresponding wave vector $| u_k \rangle$, and we calculate the \textit{Berry (Zak) Phase \cite{zak1989berry}} which represents the geometrical phase of the system through a closed loop as
\begin{equation}
\gamma_n(C)= -\text{Im} \sum_{k=0}^{M} \langle u_k | \partial_k u_k \rangle,\label{eq:berry2}
\end{equation}
\begin{figure}[ht]
 \centering
    \includegraphics[width=\columnwidth]{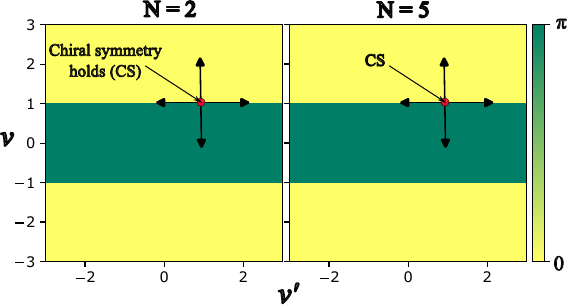}
 \caption{Berry phase in the space of free parameters of the periodic $N,i$-super-SSH model for Case I with $w=1.0$, $v'=w'$ and for different values of $N$.}
 \label{BP1}
\end{figure}
\noindent
where the sum travels a closed path in reciprocal space, \textit{i.e.}, $| u_0 \rangle = | u_M \rangle$. The $\mathbb{Z}_2$ underlying topological structure of the model allows 
to distinguish the topologically trivial $(\gamma_n(C)~=~0)$ and non-trivial $(\gamma_n(C)=\pi)$ phases of the system throughout the space of free parameters in each case. 

We also explore the spatial wave function localization of the topological zero mode state for a chain of finite length, namely, the edge states of the system, considering different sizes of the chain $n$, supercell length $N$, and impurity location within the supercell, $i$. For this purpose, we compute $|\psi_0(i)|^2$ and observe its distribution along the location $i$.


\subsection[Case I]{Case I}

In this case, we fix the parameter $w=1.0$ and force the condition $v'=w'$, thus the parameters $v$ and $v'$ define the parameter space. As can be observed in Fig.~\ref{BP1}, from the behavior of the Berry phase we notice that the trivial or non-trivial  topological phase structure of the system is not modified from the original SSH case~\cite{Batra2020,2016LNP...919.....A, 10.1088/978-1-68174-517-6}, in which the topological non-trivial phase is achieved so long as $v<w$. 
\begin{figure}[ht]
    \centering
    \includegraphics[width=\columnwidth]{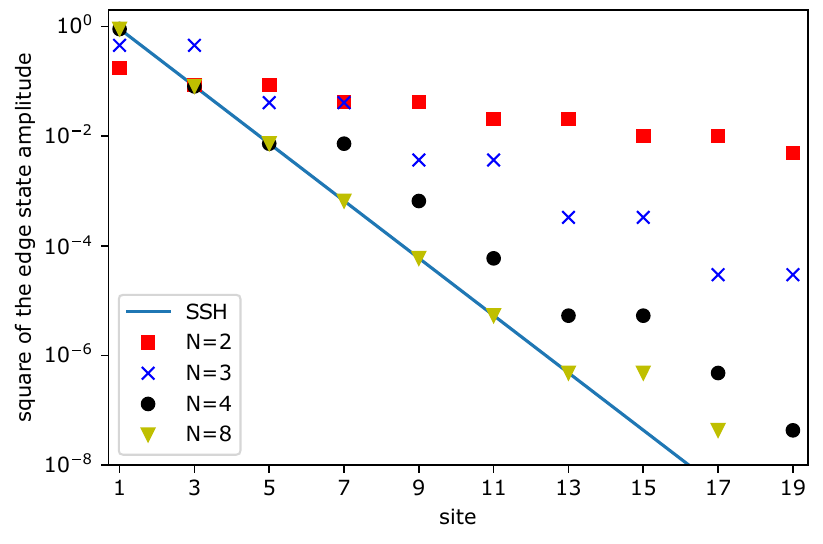}
    \caption{Edge state localization of a finite chain of the $N,i$-super-SSH model for the case $v=0.3$, $w=1.0$, $v'=w'=0.5$ for different $N$ values with the impurity in the position $i=N-1$.}
    \label{fig:loc_nvar}
\end{figure}
We can see that this condition is maintained independently of the supercell length, $N$. 
\begin{figure}[ht]
 \centering
    \includegraphics[width=0.49\textwidth]{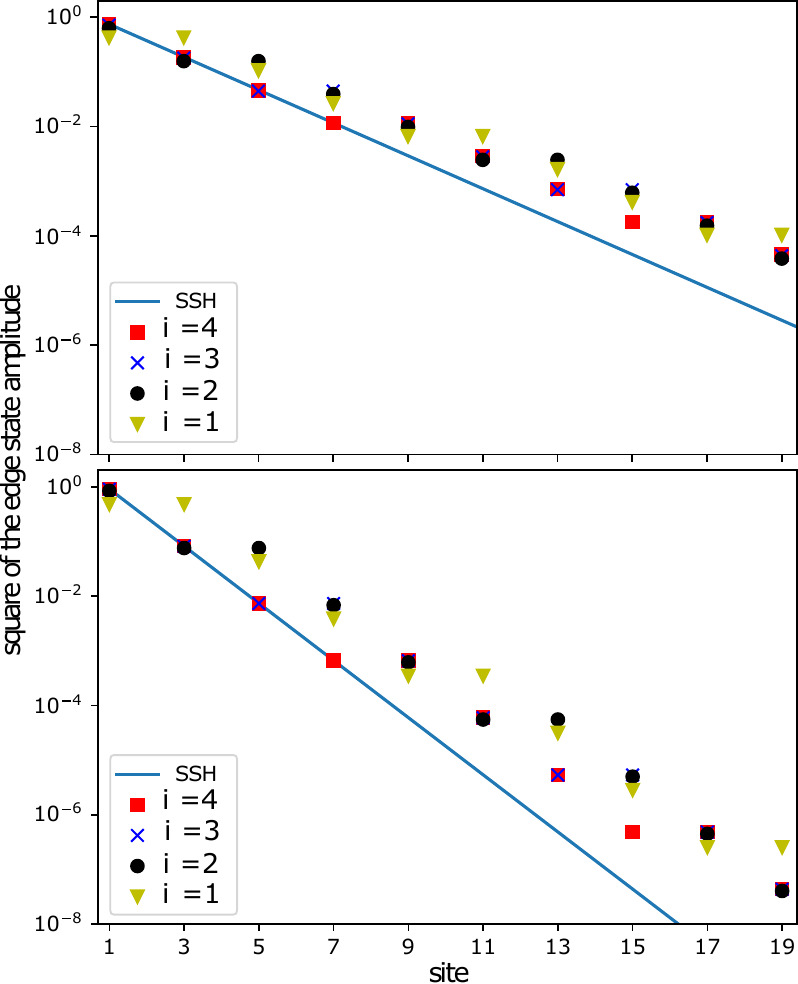}
 \caption{Edge state localization of the $N,i$-super-SSH model for the case $w=1.0$ and (a) $v=0.3$, $v'=w'=0.5$ and (b) $v=0.5$, $v'=w'=0.3$, varying the position of the impurity, $i$, for $N=4$.}
 \label{Loc}
\end{figure}
Note also that in the periodic case, the position of the impurity within the supercell is irrelevant.

 Focusing now on the localization of the zero mode edge state, we notice that the length of a finite chain composed by the concatenation of $n$ supercells is unimportant for sufficiently long chains with $n\geq 10$.  Graphics for these long chains where suppressed for space considerations in Fig~\ref{fig:loc_nvar}. For the analysis, we work with $n=12$. 
 On the other hand, as can be seen in the same Figure, the supercell length $N$ affects the edge state localization in a way that, as $N$ increases, such state resembles more and more the zero mode of the original model already in the case of $N=8$. This is completely expected since the impurity effects {\em dilute} with larger $N$, that is, the impurity density reduces for larger supercells. From this observations,  we fix  $n=12$ and $N=4$ in what follows, namely, we consider chains with $48$ sites.

Now,  from Fig.~\ref{Loc}, we observe that the edge state localization is weakened compared to the original SSH model, so that as long as the condition $v<w$ is fulfilled; the parameter $v'$ is not relevant in this case. On the contrary, the position of the impurity in the supercell is relevant, making the localization stronger for more to the right positions and weaker for more to the left positions. So that, the impurity in this case is promoting edge localization the closer it is to the right edge.

\subsection[Case II]{Case II}
In this case, we have the two benchmark SSH scenarios: ($w=1.0$; $v=0.5$)  which corresponds to a non-trivial topological phase in the SSH model and ($w=0.5$; $v=1.0$) for a trivial topological phase. The free-parameters are $v'$ and $w'$. 
\begin{figure}[ht]
 \centering
    \includegraphics[width=0.49\textwidth]{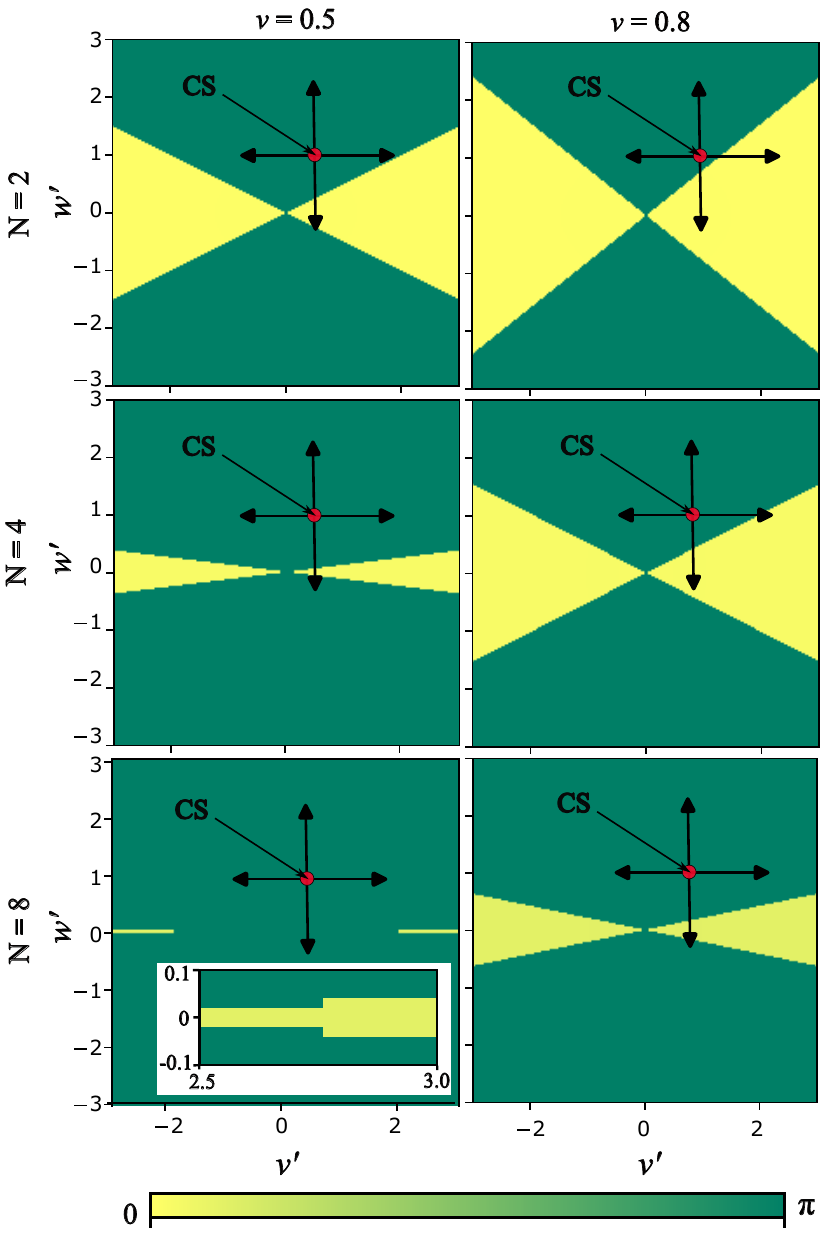}
    \caption{Berry phase in the space of free parameters ($v'$, $w'$) of the $N,i$-super-SSH model for the case $v<w$ ($v=0.5, 0.8$ and $w=1.0$ were taken) fixed for different $N$ values. }
 \label{BP2}
\end{figure}
\begin{figure}[ht]
 \centering
    \includegraphics[width=0.49\textwidth]{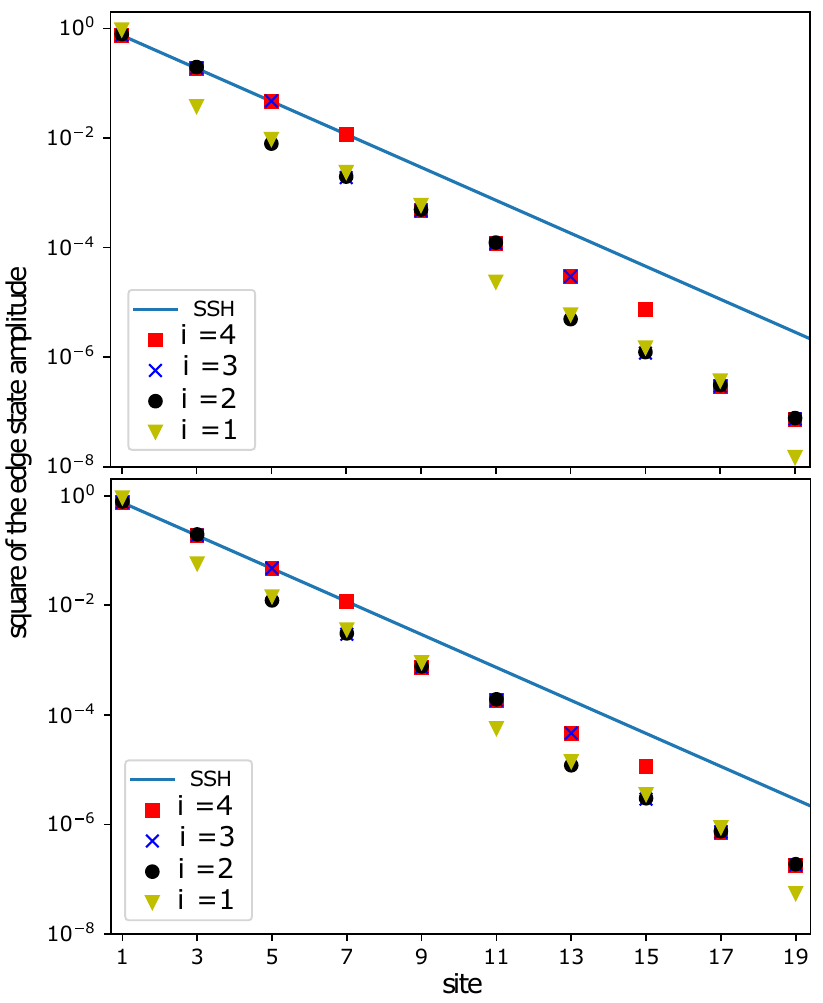}
 \caption{Edge state localization of the $N,i$-super-SSH model for the case $v<w$ ($v=0.5$ and $w=1.0$ were taken) fixed varying the free parameters $v'$,$w'$ and the position of the impurity for the case $N=4$. (Top) $v'=0.4$, $w'=2.0$; (Bottom) $v'=0.2$, $w'=0.8$.}
 \label{Loc1}
\end{figure}

As can be observed in Fig.~\ref{BP2}, for the first scenario, it is possible to have both non-trivial and trivial topological phases depending on the values of $v'$ and $w'$. These phases are separated by a linear boundary in parameter space. 
This is a counter-intuitive behavior as compared to the original SSH model. Note that for $N=2$, there are many configurations in which it is possible to switch from the non-trivial to the trivial phase, and as the supercell length increases, the possibilities of swapping are reduced. For example, for $N=8$, very large $v'$ and very small $v'$ hoppings are needed to flip between phases.
\begin{figure}[ht]
 \centering
    \includegraphics[width=\columnwidth]{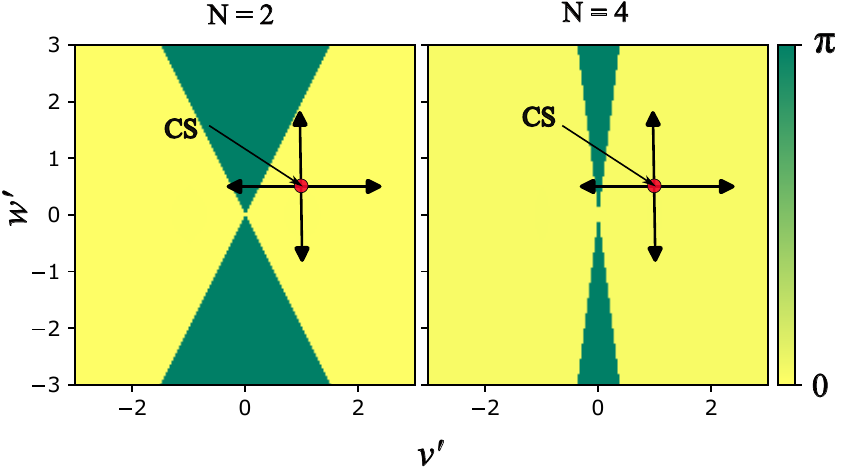}
 \caption{Berry phase in the space of free parameters of the $N,i$-super-SSH model for the case $v>w$ ($v=1.0$ and $w=0.5$ were taken) fixed for different $N$ values.}
 \label{BP3}
\end{figure}
Additionally, in Fig.~\ref{Loc1} we can observe the possibility of reaching a more pronounced edge state localization as compared to the original SSH zero mode with the appropriate tuning  of the parameters $v'$ and $w'$. For example in case of Fig.~\ref{Loc1} (top) it is accomplished with large values of $w'$, whereas in the case of Fig.~\ref{Loc1} (bottom) small values of $v'$ are the responsible of this unexpected behavior. Different from Case I, here the edge localization is reinforced as the impurity locates more toward the left in the supercell, since, for example, for the impurity position $i=1$, we obtain the largest localization.

A similar discussion follows in the scenario  $w=0.5$; $v=1.0$, as observed in Fig.~\ref{BP3}. It is well-known that for these values, the original SSH model exhibits a topologically trivial character. Surprisingly, in our model we see that it is possible to tune the system to a non-trivial topological phase by an appropriate choice of the impurity hopping. Thus, the possibility of switching to the non-trivial phase reduces as $N$ becomes larger and the impurity dilutes. 
\begin{figure}[ht]
 \centering
    \includegraphics[width=0.49\textwidth]{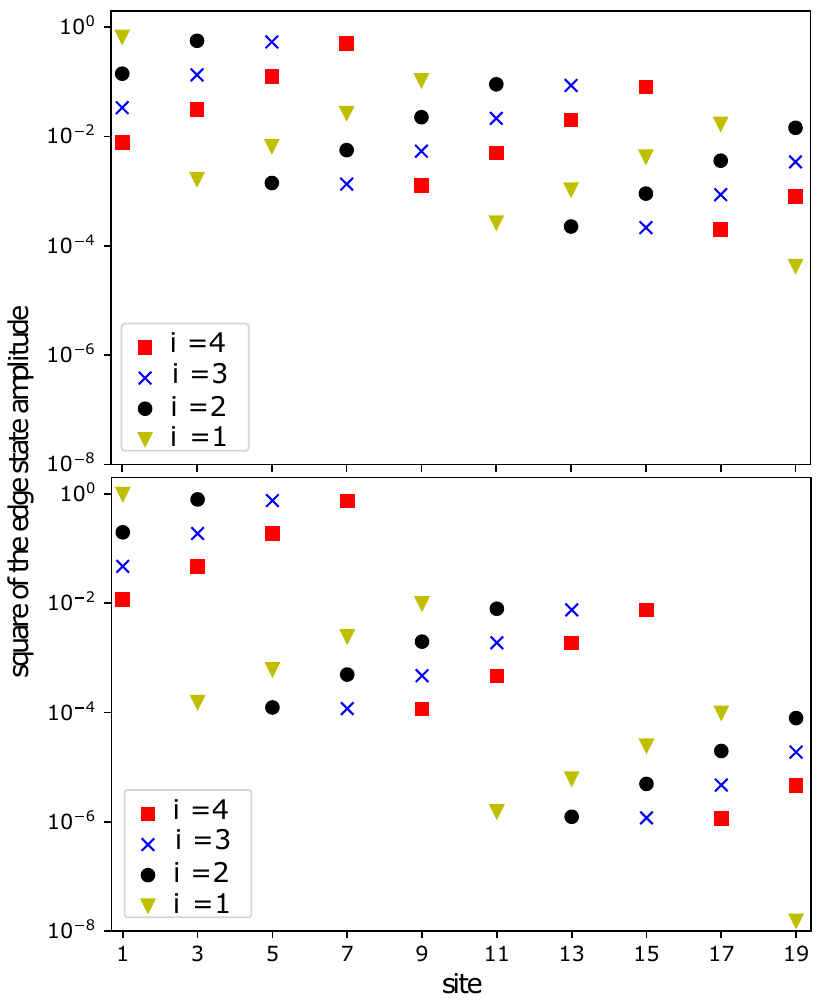}
 \caption{Edge state localization of the $N,i$-super-SSH model for the case $v>w$ ($v=1.0$ and $w=0.5$ were taken) fixed varying the free parameters $v'$,$w'$ and the position of the impurity for the case $N=4$. (Top) $v'=0.05$, $w'=1.0$; (Bottom) $v'=0.05$, $w'=4.0$.}
 \label{Loc2}
\end{figure}
\begin{figure*}[ht]
 \centering
    \includegraphics[width=0.8\textwidth]{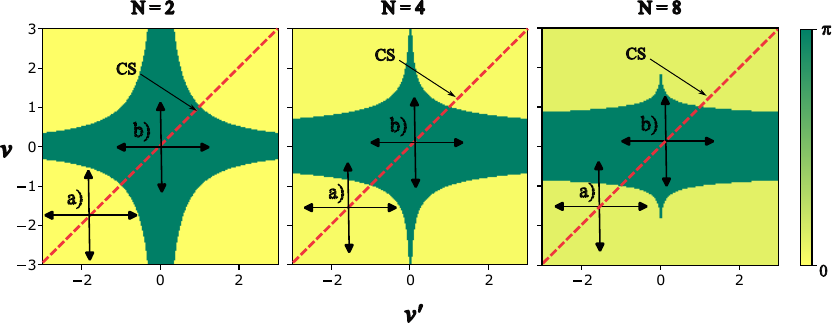}
 \caption{Berry phase in the space of free parameters of the $N,i$-super-SSH model for the case $w=w'=1.0$ for different $N$ values.}
 \label{BP4}
\end{figure*}
Notice also that just as in the previous Case, a linear interface between topological and non-topological phases emerges.

Figure~\ref{Loc2} shows the interesting and counter-intuitive behavior of the edge states localization in this case. From Fig.~\ref{BP3} we know that we require small values of $v'$, and as $w'$ is greater we find that the localization is reinforced. In addition, the position of the impurity again plays a key role since the edge state concentrates around its location, moving from previous sites and becoming more pronounced. Then, it present a jump to a less localized form in the next supercell, where then it becomes mildly localized until eventually it becomes more localized and it jumps again. In this sense, the localization becomes more pronounced for impurities placed more toward the left in the supercell. 

\subsection[Case III]{Case III}
Next we replicate the conditions of Case I with a different restriction over $w'$, namely $w=w'=1.0$. As can be observed from Fig.~\ref{BP4}, the region in parameter space where the topological phase is found gets important modifications, relaxing the topological condition $v<w$ of the ordinary model. Furthermore, the topologically trivial and non-trivial phases are separated by a more intricated shape of the boundary. Note that for $N=2$, the parameter space is symmetric under the exchange $v \leftrightarrow v'$.  On the other hand, as the supercell length increases,  the impact  of the parameter $v$ increases whereas the impact of $v'$  reduces, and for large values of $N$, the boundaries of the topologically trivial and non-trivial phases of the model tends to be as in the original SSH model (Fig.~\ref{BP1}). This makes sense because while $N$ increases, the impurity density decreases and the effects of the impurity hoppings reduces. 
\begin{figure}[ht]
 \centering
    \includegraphics[width=0.49\textwidth]{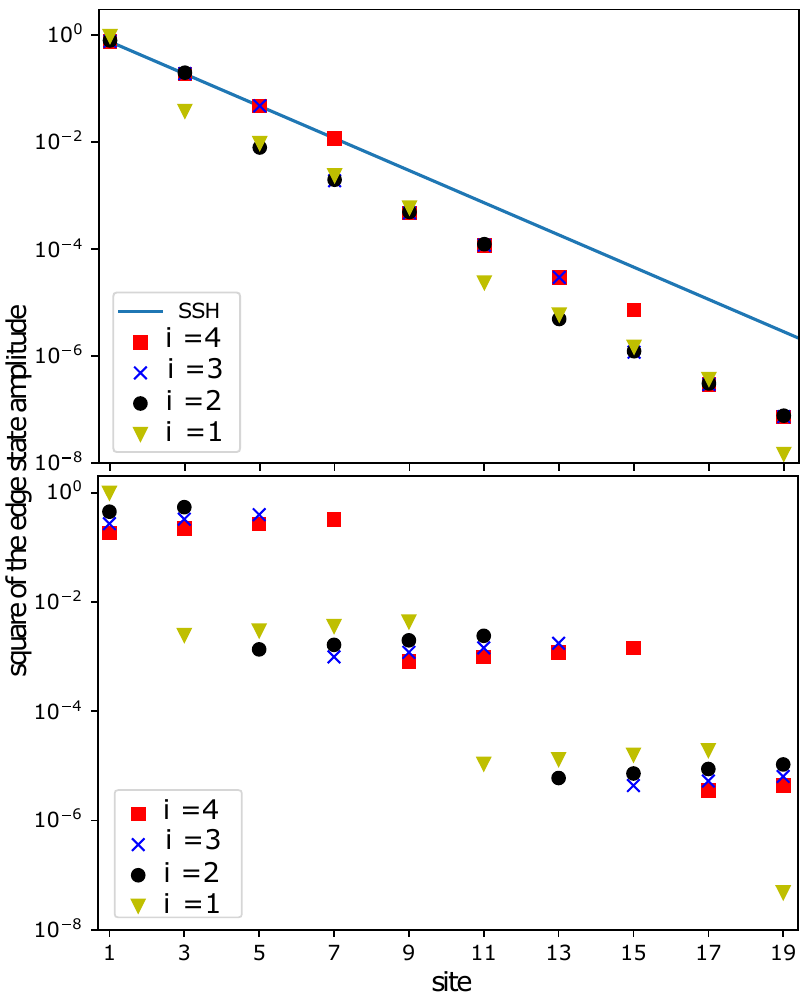}
 \caption{Edge state localization of the $N,i$-super-SSH model for the case $w=w'=1.0$ varying the free parameters $v$,$v'$ and the position of the impurity for the case $N=4$. (Top) $v=0.5$, $v'=0.2$; (Bottom) $v=1.1$, $v'=0.05$.}
 \label{Loc3}
\end{figure}

Regarding the edge state localization, in Fig.~\ref{Loc3} we observe the same behavior as in Case II. In Fig.~\ref{Loc3}(top), it can be seen that through the tuning of impurity hopping parameter, it is possible to reinforce the localization as compared to the original case. 
On the other hand, Fig.~\ref{Loc3}(bottom) shows the possibility of reaching to a non-typical non-trivial topological phase for values $v>w$. In both cases, the more to toward the left the impurity is located within the supercell, the more pronounced the localization becomes.

\section{Topolectrical circuit calculations} \refstepcounter{section}\label{sec:circuit}

\begin{figure*}[ht!]
   \centering      
   \includegraphics[width=0.8\textwidth]{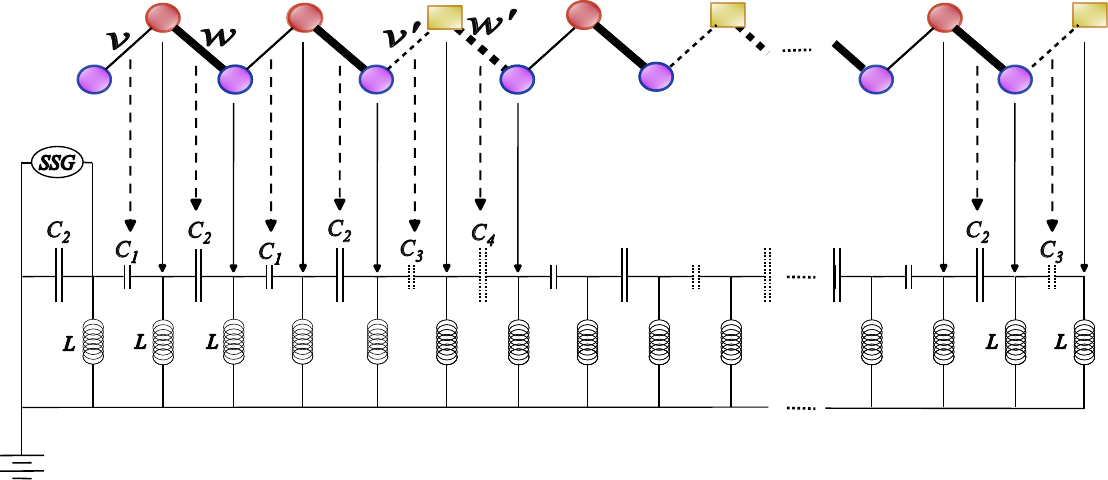}
   \caption{Topolectric circuit set up of the $N,i$-super-SSH model for $N=4$ with the impurity placed in position $i=4$.}
    \label{topoel1}
\end{figure*}
In this Section we explore again the topological phase structure of the $(N,i)$-super-SSH model from an analogue topolectrical circuit. For a self-contained discussion of the subject, in the Appendix we present a brief introduction to the framework  of topolectrical circuits and the analogy of their effective Hamiltonian with the Hamiltonian of a TB approach~\cite{PhysRevB.100.165419,doi:10.1142/S2529732520970020,RevModPhys.93.025005,Lee2018,ZHAO2018289,PhysRevResearch.3.023056}. The connection is established from the observation that the role of the hopping amplitudes in the TB approach are mapped to the capacitances  $C_i$ of capacitors in the network, whereas the ``orbital'' sites are considered as the nodes in which two capacitors and an inductor intersect. In order to get trace of the topological edge states, we look for a resonance in the impedance of the circuit with a sinusoidal signal probe. We further study the localization of the states by the voltage drops in the nodes. We simulate our topolectrical circuit within the Pyspice package~\cite{PySpice}. The topolectrical circuits in this case have the network structure shown  in Fig.~\ref{topoel1}. For all the next calculations we computed the circuit's response to an sinusoidal input voltage current with amplitude of $V_0=10 {\rm V}$, frequency $f=100 {\rm kHz}$, and inductance value $L= 10\mu {\rm H}$.

\subsection[Case I]{Case I}

\begin{figure*}[ht]
 \centering
    \includegraphics[width=0.9\textwidth]{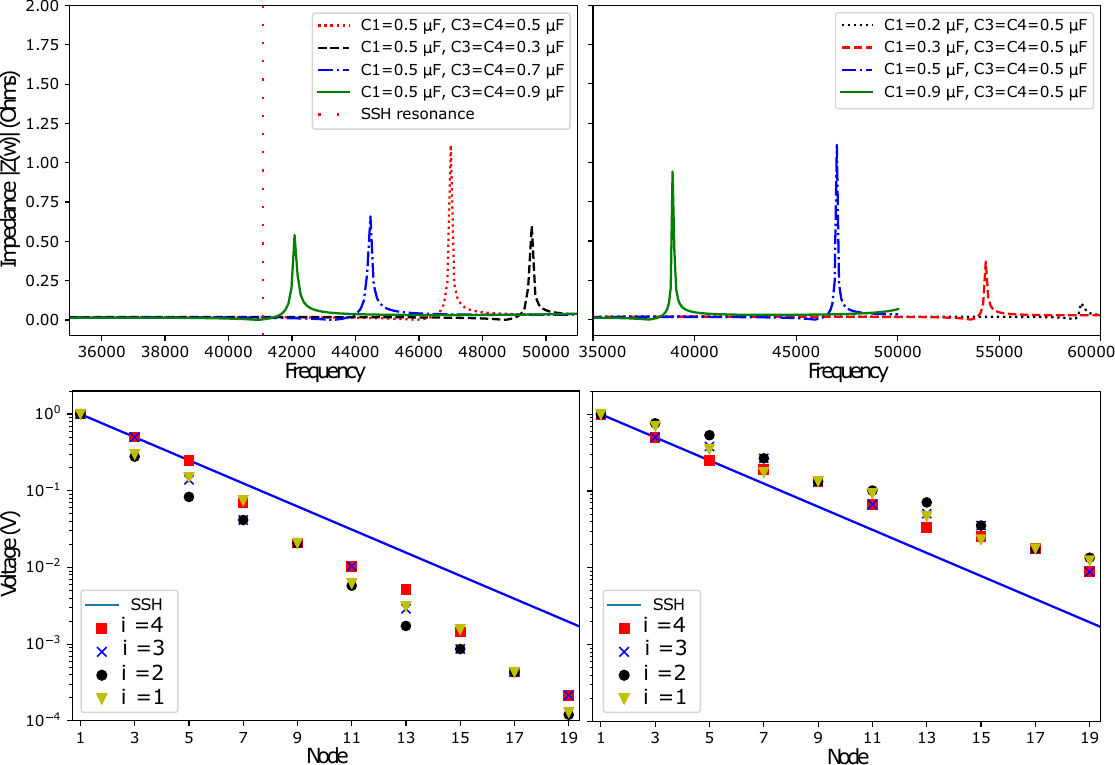}
 \caption{(Top) Impedance resonance $Z_r(\omega)$ for the values $C_2 = 1.0\mu {\rm F}$ and (left) $C_1 = 0.5\mu {\rm F}$, for different values of $C_3 = C_4$; (right) different values of $C_1$ with $C_3 = C_4 = 0.5\mu {\rm F}$. (Bottom) Edge state localization of the $N,i$-super-SSH topolectrical circuit for the case $C_2=1.0\mu {\rm F}$, (left) $C_1 = 0.5\mu {\rm F}$, $C_3 = C_4 = 0.3\mu {\rm F}$ and (right) $C_1 = 0.5\mu {\rm F}$, $C_3 = C_4 = 0.7\mu {\rm F}$, varying the position of the "impurity" (capacitors) for $N=4$.}
 \label{circ1}
\end{figure*}

In analogy with Case I of the TB calculation, we consider  $C_1$ and $C_3$ as free parameters, fixing $C_2=1.0\mu {\rm F}$ and with the condition $C_3=C_4$, \textit{i.e.} the same capacitance value to both nodes connected from the ``impurity capacitor''. In Fig.~\ref{circ1}(top) we depict the expected resonances of the impedance ($Z_r$) of the circuit as a function of the frequency of the signal probe. Such resonance appears only when the systems is found in the non-trivial topological phase~\cite{PhysRevB.100.165419,doi:10.1142/S2529732520970020,RevModPhys.93.025005,Lee2018,ZHAO2018289,PhysRevResearch.3.023056}. Figure~\ref{circ1}(top, right) shows the behavior of the resonance under the variation of  $C_1$. 
To understand the behavior of these curves, we recall that  the resonance frequency for the original SSH model is found when 
\begin{equation}
\omega_r=\frac{1}{\sqrt{L(C_1+C_2)}}.\label{resSSH}
\end{equation}
We draw these values of  $\omega_r$ as vertical dotted lines in the Figure. We notice a deviation from these values in our model due to the ``impurity'' capacitances $C_3$ and $C_4$. We can conclude that, the value of $C_1$ slightly displaces the resonance peaks of the impedance.  
\begin{figure*}[ht]
 \centering
    \includegraphics[width=0.9\textwidth]{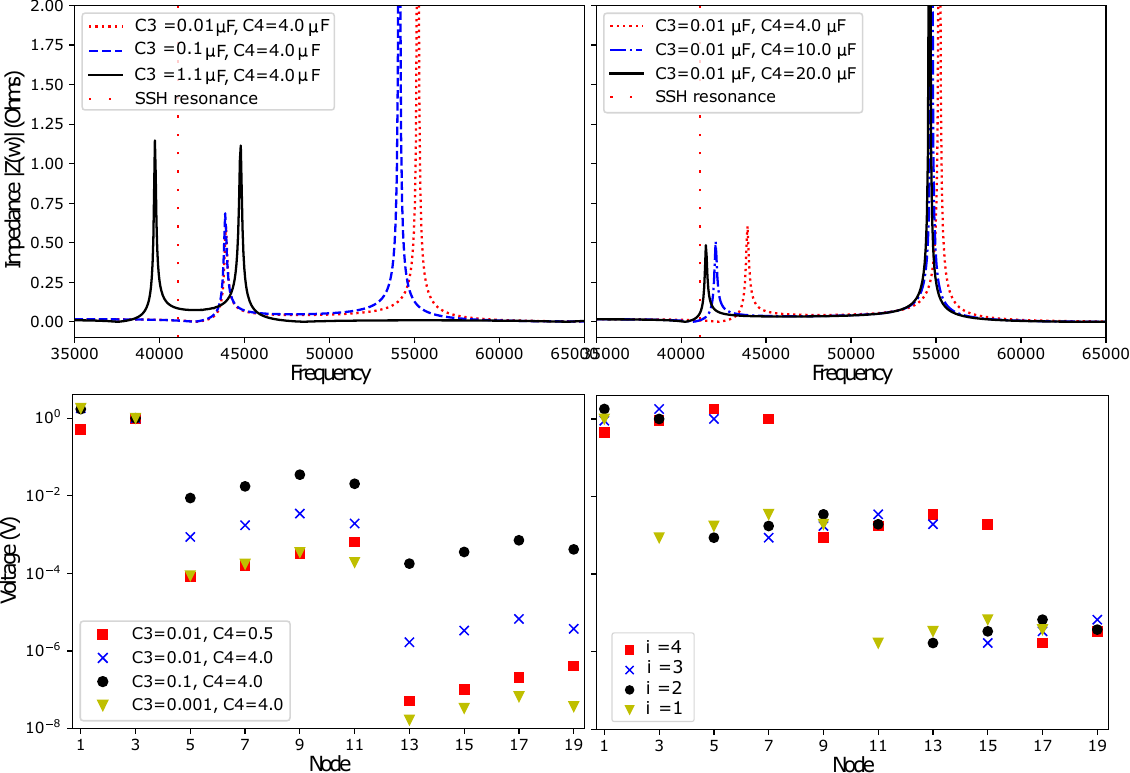}
 \caption{(Top) Impedance resonances $Z_r(\omega)$ for the values $C_2 = 1.0$ and (left) $C_1 = 0.5\mu {\rm F}$, for different values of $C_3 = C_4\mu {\rm F}$; (right) different values of $C_1$ with $C_3 = C_4 = 0.5\mu {\rm F}$. (Bottom) Edge state localization of the $N,i$-super-SSH topolectrical circuit for the case $C_2=1.0\mu {\rm F}$, (left) $C_1 = 0.5\mu {\rm F}$, $C_3 = C_4 = 0.3\mu {\rm F}$ and (right) $C_1 = 0.5\mu {\rm F}$, $C_3 = C_4 = 0.7\mu {\rm F}$, varying the position of the "impurity" (capacitors) for $N=4$.}
 \label{circ2}
\end{figure*}
On the other hand, analyzing Fig.~\ref{circ1}(top, left), as we fixed $C_1$ all the resonance peaks share essentially the same impedance $Z(\omega)$ so that we can easily detect that the effect of the variation of the $C_3=C_4$ is to displace the resonance to the left of the frequency spectra as this parameter grows, approaching to the value of the original SSH resonance frequency, $\omega_r$.  

In close analogy with the wave function localization of the edge states, the voltage in the topological circuit is higher at the edge of the circuit. The voltage localization is shown in Fig.~\ref{circ1}(bottom), where we can observe that for values $C_3<C_1$ (bottom, left) the localization is reinforced with respect to the topolectrical SSH circuit shown in the blue line, while in the opposite case (bottom, right), the localization is weakened. The position of the ``impurity'' in the circuit also plays a role, reproducing the same behavior as in the TB calculations where the more to the right the impurity is located, the more pronounced the localization becomes. We must also mention that as in the TB calculation, the effect of the circuit length is irrelevant for long enough chains of circuit elements and that the effect of the supercell length is to dilute impurity effects as $N$ gets large. 

\subsection[Case II]{Case II}

In this case we translate the conventions of Case II of the  TB calculation, and study the well-known topologically nontrivial ($C_1<C_2$) and trivial ($C_1>C_2$) cases under the variation of the free parameters $C_3$ and $C_4$. We compute the impedance of the system and identify its resonant value $Z_r$ upon varying the values of $C_3$ and $C_4$. We are mostly interested in the non-intuitive cases in which a non-trivial topological phase is found even though $C_1>C_2$. In Fig.~\ref{circ2} we present the resonances of the impedance and the voltage localization for $C_1=1.1\mu{\rm F}$, $C_2=1.0\mu{\rm F}$, varying $C_3$ and $C_4$. From Fig.~\ref{circ2}(top), we observe the appearance of a second resonance peak which contributes to voltage localization. It has been shown that Fano resonances \cite{fano1961effects} appear as a dispersion mechanism in the circuits \cite{lv2016analysis, 10.1007/978-3-030-16770-7_14, molina2022fano}, and that is the responsible of the asymmetry of the resonance peaks as well as the secondary peak appearance. The effect of $C_4$ is to slightly displace resonance peaks to the left as it grows. On the other hand, the effect of $C_3$ is to displace the resonances peaks to the left as it grows.

Voltage localization in the circuit is depicted in Fig.~\ref{circ2}(bottom). We observe on the one hand the same behavior regarding the impurity position as in the TB approach, that is, a staggered localization around the impurity with jumps every supercell, and a reinforcing localization at the edge as the impurity is located more to the left. Also, keeping fixed the impurity position ($i=2$), localization becomes more pronounced as $C_3$ and $C_4$ decrease (c.f. Fig.~\ref{Loc2}).
\subsection[Case III]{Case III}
\begin{figure}[ht]
 \centering
    \includegraphics[width=0.49\textwidth]{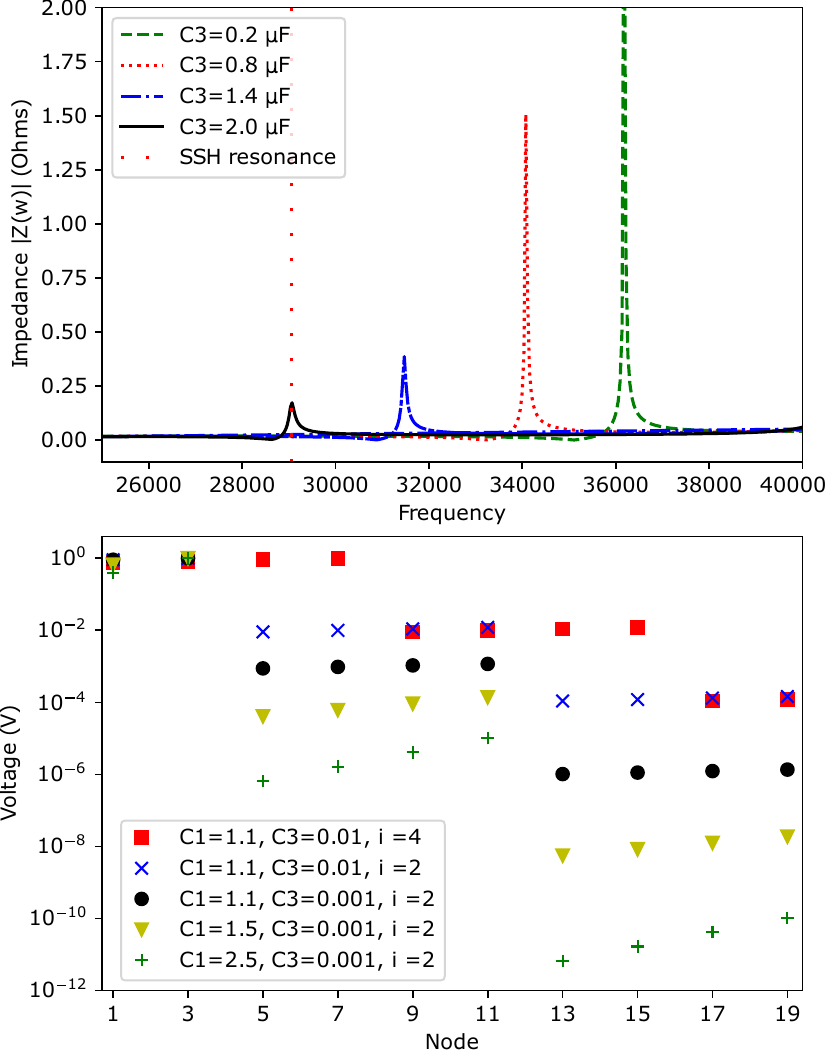}
 \caption{(Top) Impedance resonances $Z_r(\omega)$ for the values $C_2 = C_4 1.0\mu {\rm F}$ and $C_1 = 2.0\mu {\rm F}$, for different values of $C_3$. (Bottom) Edge state localization of the $N,i$-super-SSH topolectrical circuit for the case $C_2 = C_4=1.0\mu {\rm F}$ varying the position of the "impurity" (capacitors) for $N=4$.}
 \label{circ3}
\end{figure}

For this Case, which adopts the conventions of Case III of the TB calculation with $C_2=C_4=1.0\mu {\rm F}$, we focus our attention into the non-intuitive region in which even for $C_1>C_2$ (notice that $C_1=2.0\mu{\rm F}$ is taken in Fig.~\ref{circ3}(top)),  a non-trivial topological phase develops. Thus, the only free parameter is actually $C_3$. The resonance peak of the impedance for the corresponding topolectrical circuit is plotted in Fig.~\ref{circ3}(top) for different values of $C_3$. 

Finally, in Fig.~\ref{circ3}(bottom) we show the voltage drops in the different nodes of the topolectrical circuit for different values of $C_1>1\mu{\rm F}$, $C_3$ and the impurity position, $i$. 
We observe the same behavior as with the TB approach: A staggered localization around the impurity with jumps every supercell, a reinforcing localization at the edge as the impurity places more to the left, and the greater $C_1$ and the smaller $C_3$ the more pronounced the localization becomes (compare with Fig.~\ref{Loc3}).

\section{Conclusions} \refstepcounter{section}\label{sec:conclusions}
In this paper we have performed an study of the effects of embedded impurity superlattices in the SSH model over electronic and topological properties of the system. We considered impurity effects through the modification of the hopping amplitudes between these impurities and their nearest neighbours in the array. Three different cases of these modified hoppings were considered: the Case I in which $w=1.0$, $v'=w'$; Case II, where $v,w$ where fixed (=1.0, 0.5, respectively, and vice versa); and Case III, in which $w=1.0$, $w'=w$. We used two frameworks to explore the proteries of the system: the TB approach and a topolectrical circuit analogy.

Using the TB approach, for the Case I, we obtain no modification of the topological boundaries of the different phases of the model as compared to the original SSH model, indicating that this kind of impurities do not impact the topological condition for finding the topological phase of the model. Moreover,  the position of the impurity within the supercell plays an important role in strengthening the localization of the edge-states, specially as the impurity location is more to the right edge of the supercell. For Cases II and III, non-intuitive, interesting boundaries of the topologically different phases were obtained: a linear boundary that separates topologically non-trivial and trivial phases, although the ordinary condition $v<w$ or $v>w$ is no longer required, and a non-linear-boundary between these phases. Effects are more notorious for small $N$, recovering the usual SSH model behavior for large $N$, as the impurity effects get diluted. In these cases, the position of the impurity within the supercell is important in the localization of the edge-states just as in the first case, but now reinforcing localization as the impurity is placed more to the left.

Using the topolectrical circuit analogue of the system, the expected resonances in the impedance of the circuit with a sinusoidal signal probe where found in each case, deviating from the resonant frequency $\omega_r$ of the original SSH model because of the presence of the impurity. The free parameters in each case are relevant in the position  of the resonance peaks, while the same behavior in the voltage localization was realized as in the TB approach calculation. In all cases, the effect of the system length, $n$, is irrelevant for sufficiently long systems, with $n\leq10$, while the supercell length, $N$, reduces the effect of the impurities in the systems as $N$ gets larger, because of the impurity {\em dilution} in the system. The topolectrical circuit analogue of these models confirmed the results obtained in the TB approach. Additionally, within the numerical simulation of the circuit, the existence of other resonances was shown, related to Fano resonance in the system, and  perhaps these are responsible of small differences between models. Even though, we conclude that topological circuits are as useful for studying electric and topological features of materials as TB calculations.

A general remark is that the results of this investigation provide evidence for the possibility of relaxing the conditions for the hopping amplitudes to distinguish between topologically trivial and non-trivial phases in the model and to provide a more tight localization of the edge-states by means of embedded impurity superlattices. Impurities in non-topological materials could be an important tool to provide them with topological features. Furthermore, impurities could be of interest for technological applications by tightening the localization of the wave vectors, that is, improving the surface conduction in these materials.

\section{Acknowledgements}
JCPP and AR would like to thank support from CONAHCyT under grant FORDECYT-\newline PRONACES/61533/2020. AR acknowledges enlightening discussions with Prof. Enrique Muñoz.

\section*{Topolectrical circuits}
Electrical circuits 
are described by Kirchoff's laws. Following~\cite{PhysRevResearch.3.023056}, we start from Kirchoff's current law
\begin{equation}\label{KL}
    I_a = \sum_i c_{ai}(V_a-V_i) + w_a V_A,
\end{equation}
which establishes that the input current in a node, $I_a$, is equal to the current flowing out this node to other node $i$ linked by a conductance $c_{ai}$ plus the the current flowing to ground with impedance $w^{-1}_a$. Eq.~(\ref{KL}) can be expressed in matrix form as 
\begin{equation}\label{matrix}
    \Vec{I} = (N+W)\Vec{V} = J \Vec{V},
\end{equation}
where $N$ is the circuit~\textit{Laplacian} depending on the conductances, $W$ is a diagonal matrix depending on the circuit's grounding, and $J=N+W$ is the grounded Laplacian, also called admittance matrix. Additionally, the Laplacian can be decomposed as $N=D-C$, with $C$ the adjacency or conductances matrix and $D$ a diagonal matrix representing the list of total currents out of each node. For LC circuits, the driving voltage frequency $\omega$ is the cornerstone. In this sense, it is convenient to Fourier transform Eq.~(\ref{matrix}), obtaining~\cite{PhysRevResearch.3.023056}
\begin{equation}
    I_a(\omega) = \sum_b J_{ab}(\omega) V_b(\omega).
\end{equation}
The admittance matrix now reads
\begin{equation}
    J_{ab}(\omega) = i\omega [N_{ab}(\omega)+\delta_{ab} W_a(\omega)],
\end{equation}
where
\begin{equation}
    N_{ab}(\omega)= -C_{ab}+ \frac{1}{\omega^2 L_{ab}},
\end{equation}
and
\begin{equation}
    W_{a}(\omega)= C_{a}- \frac{1}{\omega^2 L_{a}}-\sum_c N_{ac},
\end{equation}
where $C_{ab}$ and $L_{ab}$ are the capacitance and inductance between nodes $a$ and $b$, respectively, and $C_{a}$ and $L_{a}$ are the capacitance and inductance between nodes $a$ and the ground.
Most commonly, circuits are often studied by its voltage response to an applied current. This can be done through the corresponding impedance, $Z$. The two-point impedance is calculated as \cite{cserti2011uniform}
\begin{equation}\label{imp}
    Z_{ab} = \frac{(V_a-V_b)}{I},
\end{equation}
where $I=|I_{ab}|$. Expression in Eq.~(\ref{imp}) is used to calculate the impedance of different circuits throughout this paper. Expressing the potentials in terms of the input current, that is inverting Eq.~(\ref{matrix}), we use the regularized circuit Green's function, 
\begin{equation}
G= \sum_{j_n\neq 0} \frac{1}{j_n} \psi_n \psi_n^{\dagger},
\end{equation}
where $j_n$ denotes the admittance eigenvalues of $\vec{J}$ and $\psi_n$ is the corresponding $n-$dimensional eigenvector matrix. Notice that the admittance is defines as $Y=Z^{-1}$. Thus, it is possible to express the impedance as
\begin{equation}\label{resonance}
    Z_{ab} = \sum_{j_n\neq 0} \frac{|\psi_{na} -\psi_{nb}|^2}{j_n},
\end{equation}
that is, the sum of the squared modulus of the difference between of the $n-$th admittance eigenmode divided by the admittance eigenvalue. As can be seen from Eq.~(\ref{resonance}), the impedance becomes larger, namely, exhibits a resonance, if the admittance eigenmodes are well-localized at one region with small admittance eigenvalue $j_n$. This is the case of topological boundary resonances in topolectrical circuits, where there exists a large density of protected boundary modes with $j_n\approx 0$.  

To make the correspondence between topolectrical circuits and Tight-Binding approaches, we should note that we are dealing with two eigenvalue equations:
\begin{align}
    \text{Circuits:\ }&\qquad J\vec{V} \equiv j\vec{V}= \vec{I},\\
    \text{QM:\ }&\qquad H\vec{\psi} = E\vec{\psi}.
\end{align}
If we relate the nodes in the topolectrical circuit to lattice sites, it is possible to establish the relation $J_{ab}(\omega) = i\omega H_{ab}(\omega)$, with $H(\omega)$ the Hamiltonian of the TB approach~\cite{PhysRevResearch.3.023056}.

\subsection[Topolectrical circuit of the (N,i)-super-SSH model.]{Topolectrical circuit of the \texorpdfstring{$(N,i)$}--super-SSH model.}

The topolectrical analogue of the $(AB)^{N-1}-AC-SSH$ TB model in accomplished by considering the lattice sites as the nodes $1,2, ...$ of the circuit in Fig.~\ref{topoel1}. The hopping amlitudes between sites $A$ and $B$ and vice versa are set as capacitance values $C_1$ and $C_2$, respectively. An impurity is represented by a change in the capacitance linking its site with its nearest neighbours sites by the values of $C_3$ and $C_4$ (Fig.~\ref{topoel1}).

The two-point impedance is calculated using Eq.~(\ref{imp}) by measuring the voltage in node 1. Additionally, the voltage localization is computed by measuring the voltage in each node. 

For comparative reasons, it is be useful to establish some characteristics of the original topolectrical SSH circuit ($C_3=C_1$ and $C_4=C_2$)~\cite{PhysRevResearch.3.023056}. First, when the condition $C_1<C_2$ fulfills, a topological boundary mode exists, leading to a drastic increase in impedance (resonance) which presents in the resonant frequency \eqref{resSSH}. Also, applying an alternate current probe generates potential differences between the plates of each capacitor, $V_1$ and $V_2$ for capacitors $C_1$ and $C_2$, respectively, which indeed oscillate in anti-phase. When $C_1<C_2$, the potential configuration on the nodes (mid-gap mode) is given by $\psi_0(n) \alpha (1,0,-t,0,t^2,0-t^3,0,...,((-t]^n,0))$, with $t=C_1/C_2$. In the opposite case in which $C_1>C_2$ ($t>1$), there exists no topological boundary mode and no mid-gap node is found. Finally, In terms of the grounded Laplacian, the system with periodic boundary conditions is described by 
\begin{widetext}
\begin{equation}
J_{SSH}(k_x) = i\omega \left(C_1+C_2-\frac{1}{\omega^2 L}\right) I - i\omega\left[ (C_1+C_2 \cos k_x)\sigma_x + C_2 \sin k_x \sigma_y \right].
\end{equation}
\end{widetext}


\newpage
\section*{References}

\bibliography{biblio}

\begin{thebibliography}{10}

\bibitem{von1986quantized}
Klaus Von~Klitzing.
\newblock The quantized hall effect.
\newblock {\em Reviews of Modern Physics}, 58(3):519, 1986.

\bibitem{klitzing1980new}
K~v Klitzing, Gerhard Dorda, and Michael Pepper.
\newblock New method for high-accuracy determination of the fine-structure
  constant based on quantized hall resistance.
\newblock {\em Physical review letters}, 45(6):494, 1980.

\bibitem{stormer1999fractional}
Horst~L Stormer, Daniel~C Tsui, and Arthur~C Gossard.
\newblock The fractional quantum hall effect.
\newblock {\em Reviews of Modern Physics}, 71(2):S298, 1999.

\bibitem{kane2005z}
Charles~L Kane and Eugene~J Mele.
\newblock Z 2 topological order and the quantum spin hall effect.
\newblock {\em Physical review letters}, 95(14):146802, 2005.

\bibitem{Batra2020}
Navketan Batra and Goutam Sheet.
\newblock Physics with coffee and doughnuts.
\newblock {\em Resonance}, 25(6):765--786, Jun 2020.

\bibitem{2016LNP...919.....A}
J{\'a}nos~K. {Asb{\'o}th}, L{\'a}szl{\'o} {Oroszl{\'a}ny}, and Andr{\'a}s
  {P{\'a}lyi}.
\newblock {\em {A Short Course on Topological Insulators}}, volume 919.
\newblock 2016.

\bibitem{10.1088/978-1-68174-517-6}
Panagiotis Kotetes.
\newblock {\em Topological Insulators}.
\newblock 2053-2571. Morgan Claypool Publishers, 2019.

\bibitem{periodic}
Alexei Kitaev.
\newblock {Periodic table for topological insulators and superconductors}.
\newblock {\em AIP Conference Proceedings}, 1134(1):22--30, 05 2009.

\bibitem{ssh}
W.~P. Su, J.~R. Schrieffer, and A.~J. Heeger.
\newblock Solitons in polyacetylene.
\newblock {\em Phys. Rev. Lett.}, 42:1698--1701, Jun 1979.

\bibitem{PhysRevB.22.2099}
W.~P. Su, J.~R. Schrieffer, and A.~J. Heeger.
\newblock Soliton excitations in polyacetylene.
\newblock {\em Phys. Rev. B}, 22:2099--2111, Aug 1980.

\bibitem{RevModPhys.82.3045}
M.~Z. Hasan and C.~L. Kane.
\newblock Colloquium: Topological insulators.
\newblock {\em Rev. Mod. Phys.}, 82:3045--3067, Nov 2010.

\bibitem{RevModPhys.83.1057}
Xiao-Liang Qi and Shou-Cheng Zhang.
\newblock Topological insulators and superconductors.
\newblock {\em Rev. Mod. Phys.}, 83:1057--1110, Oct 2011.

\bibitem{shankar2018topological}
R.~Shankar.
\newblock Topological insulators -- a review, 2018.

\bibitem{wang2013topological}
Lei Wang, Matthias Troyer, and Xi~Dai.
\newblock Topological charge pumping in a one-dimensional optical lattice.
\newblock {\em Physical review letters}, 111(2):026802, 2013.

\bibitem{heeger1988solitons}
Alan~J Heeger, Steven Kivelson, John~Robert Schrieffer, and W-P Su.
\newblock Solitons in conducting polymers.
\newblock {\em Reviews of Modern Physics}, 60(3):781, 1988.

\bibitem{yu1988solitons}
Lu~Yu.
\newblock {\em Solitons and polarons in conducting polymers}.
\newblock World Scientific, 1988.

\bibitem{ryu2010topological}
Shinsei Ryu, Andreas~P Schnyder, Akira Furusaki, and Andreas~WW Ludwig.
\newblock Topological insulators and superconductors: tenfold way and
  dimensional hierarchy.
\newblock {\em New Journal of Physics}, 12(6):065010, 2010.

\bibitem{peng2016experimental}
Yu-Gui Peng, Cheng-Zhi Qin, De-Gang Zhao, Ya-Xi Shen, Xiang-Yuan Xu, Ming Bao,
  Han Jia, and Xue-Feng Zhu.
\newblock Experimental demonstration of anomalous floquet topological insulator
  for sound.
\newblock {\em Nature communications}, 7(1):13368, 2016.

\bibitem{gomez2013floquet}
Alvaro G{\'o}mez-Le{\'o}n and Gloria Platero.
\newblock Floquet-bloch theory and topology in periodically driven lattices.
\newblock {\em Physical review letters}, 110(20):200403, 2013.

\bibitem{dal2015floquet}
Virginia Dal~Lago, M~Atala, and LEF~Foa Torres.
\newblock Floquet topological transitions in a driven one-dimensional
  topological insulator.
\newblock {\em Physical Review A}, 92(2):023624, 2015.

\bibitem{an2018engineering}
Fangzhao~Alex An, Eric~J Meier, and Bryce Gadway.
\newblock Engineering a flux-dependent mobility edge in disordered zigzag
  chains.
\newblock {\em Physical Review X}, 8(3):031045, 2018.

\bibitem{perez2019interplay}
Beatriz P{\'e}rez-Gonz{\'a}lez, Miguel Bello, {\'A}lvaro G{\'o}mez-Le{\'o}n,
  and Gloria Platero.
\newblock Interplay between long-range hopping and disorder in topological
  systems.
\newblock {\em Physical Review B}, 99(3):035146, 2019.

\bibitem{li2018topological}
Chun-Fang Li, Xin-Ping Li, and Lin-Cheng Wang.
\newblock Topological phases of modulated su-schrieffer-heeger chains with
  long-range interactions.
\newblock {\em Europhysics Letters}, 124(3):37003, 2018.

\bibitem{li2017topological}
C~Li, S~Lin, G~Zhang, and Z~Song.
\newblock Topological nodal points in two coupled su-schrieffer-heeger chains.
\newblock {\em Physical Review B}, 96(12):125418, 2017.

\bibitem{junemann2017exploring}
J~J{\"u}nemann, A~Piga, S-J Ran, M~Lewenstein, M~Rizzi, and Alejandro
  Berm{\'u}dez.
\newblock Exploring interacting topological insulators with ultracold atoms:
  The synthetic creutz-hubbard model.
\newblock {\em Physical Review X}, 7(3):031057, 2017.

\bibitem{sun2017quantum}
Ning Sun and Lih-King Lim.
\newblock Quantum charge pumps with topological phases in a creutz ladder.
\newblock {\em Physical Review B}, 96(3):035139, 2017.

\bibitem{li2022dirac}
Chang-An Li, Sang-Jun Choi, Song-Bo Zhang, and Bj{\"o}rn Trauzettel.
\newblock Dirac states in an inclined two-dimensional su-schrieffer-heeger
  model.
\newblock {\em Physical Review Research}, 4(2):023193, 2022.

\bibitem{xie2019topological}
Dizhou Xie, Wei Gou, Teng Xiao, Bryce Gadway, and Bo~Yan.
\newblock Topological characterizations of an extended su--schrieffer--heeger
  model.
\newblock {\em npj Quantum Information}, 5(1):55, 2019.

\bibitem{li2014topological}
Linhu Li, Zhihao Xu, and Shu Chen.
\newblock Topological phases of generalized su-schrieffer-heeger models.
\newblock {\em Physical Review B}, 89(8):085111, 2014.

\bibitem{agrawal2023cataloging}
Aayushi Agrawal and Jayendra~N Bandyopadhyay.
\newblock Cataloging topological phases of n stacked su-schrieffer-heeger
  chains by a systematic breaking of symmetries.
\newblock {\em Physical Review B}, 108(10):104101, 2023.

\bibitem{li2022topological}
Jia-Rui Li, Lian-Lian Zhang, Wei-Bin Cui, and Wei-Jiang Gong.
\newblock Topological properties in non-hermitian tetratomic
  su-schrieffer-heeger lattices.
\newblock {\em Physical Review Research}, 4(2):023009, 2022.

\bibitem{sivan2022topology}
A~Sivan and M~Orenstein.
\newblock Topology of multiple cross-linked su-schrieffer-heeger chains.
\newblock {\em Physical Review A}, 106(2):022216, 2022.

\bibitem{zak1989berry}
J~Zak.
\newblock Berry’s phase for energy bands in solids.
\newblock {\em Physical review letters}, 62(23):2747, 1989.

\bibitem{betancurocampo2023twofold}
Yonatan Betancur-Ocampo, B.~Manjarrez-Montañez, A.~M. Martínez-Argüello, and
  Rafael~A. Méndez-Sánchez.
\newblock Twofold topological phase transitions induced by
  third-nearest-neighbor interactions in 1d chains, 2023.

\bibitem{wang2018experimental}
Bo~Wang, Tian Chen, and Xiangdong Zhang.
\newblock Experimental observation of topologically protected bound states with
  vanishing chern numbers in a two-dimensional quantum walk.
\newblock {\em Physical Review Letters}, 121(10):100501, 2018.

\bibitem{chen2018observation}
Chao Chen, Xing Ding, Jian Qin, Yu~He, Yi-Han Luo, Ming-Cheng Chen, Chang Liu,
  Xi-Lin Wang, Wei-Jun Zhang, Hao Li, et~al.
\newblock Observation of topologically protected edge states in a photonic
  two-dimensional quantum walk.
\newblock {\em Physical review letters}, 121(10):100502, 2018.

\bibitem{xiao2015geometric}
Meng Xiao, Guancong Ma, Zhiyu Yang, Ping Sheng, ZQ~Zhang, and Che~Ting Chan.
\newblock Geometric phase and band inversion in periodic acoustic systems.
\newblock {\em Nature Physics}, 11(3):240--244, 2015.

\bibitem{stuhl2015visualizing}
BK~Stuhl, H-I Lu, LM~Aycock, D~Genkina, and IB~Spielman.
\newblock Visualizing edge states with an atomic bose gas in the quantum hall
  regime.
\newblock {\em Science}, 349(6255):1514--1518, 2015.

\bibitem{meier2016observation}
Eric~J Meier, Fangzhao~Alex An, and Bryce Gadway.
\newblock Observation of the topological soliton state in the
  su--schrieffer--heeger model.
\newblock {\em Nature communications}, 7(1):13986, 2016.

\bibitem{meier2018observation}
Eric~J Meier, Fangzhao~Alex An, Alexandre Dauphin, Maria Maffei, Pietro
  Massignan, Taylor~L Hughes, and Bryce Gadway.
\newblock Observation of the topological anderson insulator in disordered
  atomic wires.
\newblock {\em Science}, 362(6417):929--933, 2018.

\bibitem{cai2019experimental}
Han Cai, Jinhong Liu, Jinze Wu, Yanyan He, Shi-Yao Zhu, Jun-Xiang Zhang, and
  Da-Wei Wang.
\newblock Experimental observation of momentum-space chiral edge currents in
  room-temperature atoms.
\newblock {\em Physical Review Letters}, 122(2):023601, 2019.

\bibitem{goldman2016topological}
Nathan Goldman, Jan~C Budich, and Peter Zoller.
\newblock Topological quantum matter with ultracold gases in optical lattices.
\newblock {\em Nature Physics}, 12(7):639--645, 2016.

\bibitem{jotzu2014experimental}
Gregor Jotzu, Michael Messer, R{\'e}mi Desbuquois, Martin Lebrat, Thomas
  Uehlinger, Daniel Greif, and Tilman Esslinger.
\newblock Experimental realization of the topological haldane model with
  ultracold fermions.
\newblock {\em Nature}, 515(7526):237--240, 2014.

\bibitem{PhysRevB.100.165419}
Motohiko Ezawa.
\newblock Electric-circuit simulation of the schr\"odinger equation and
  non-hermitian quantum walks.
\newblock {\em Phys. Rev. B}, 100:165419, Oct 2019.

\bibitem{doi:10.1142/S2529732520970020}
Russell Yang~Qi Xun.
\newblock Topological circuits - a stepping stone in the topological
  revolution.
\newblock {\em Molecular Frontiers Journal}, 04(Supp01):9--14, 2020.

\bibitem{RevModPhys.93.025005}
Alexandre Blais, Arne~L. Grimsmo, S.~M. Girvin, and Andreas Wallraff.
\newblock Circuit quantum electrodynamics.
\newblock {\em Rev. Mod. Phys.}, 93:025005, May 2021.

\bibitem{Lee2018}
Ching~Hua Lee, Stefan Imhof, Christian Berger, Florian Bayer, Johannes Brehm,
  Laurens~W. Molenkamp, Tobias Kiessling, and Ronny Thomale.
\newblock Topolectrical circuits.
\newblock {\em Communications Physics}, 1(1):39, Jul 2018.

\bibitem{ZHAO2018289}
Erhai Zhao.
\newblock Topological circuits of inductors and capacitors.
\newblock {\em Annals of Physics}, 399:289--313, 2018.

\bibitem{PhysRevResearch.3.023056}
Junkai Dong, Vladimir Juri\ifmmode \check{c}\else
  \v{c}\fi{}i\ifmmode~\acute{c}\else \'{c}\fi{}, and Bitan Roy.
\newblock Topolectric circuits: Theory and construction.
\newblock {\em Phys. Rev. Res.}, 3:023056, Apr 2021.

\bibitem{xun2020topological}
Russell Yang~Qi Xun.
\newblock Topological circuits-a stepping stone in the topological revolution.
\newblock {\em Molecular Frontiers Journal}, 4(Supp01):9--14, 2020.

\bibitem{ningyuan2015time}
Jia Ningyuan, Clai Owens, Ariel Sommer, David Schuster, and Jonathan Simon.
\newblock Time-and site-resolved dynamics in a topological circuit.
\newblock {\em Physical Review X}, 5(2):021031, 2015.

\bibitem{zhu2018simulating}
Weiwei Zhu, Shanshan Hou, Yang Long, Hong Chen, and Jie Ren.
\newblock Simulating quantum spin hall effect in the topological lieb lattice
  of a linear circuit network.
\newblock {\em Physical Review B}, 97(7):075310, 2018.

\bibitem{pythTB}
phythTB.
\newblock Python tight binding, 2016.

\bibitem{PySpice}
PySpice.
\newblock Pyspice, 2021.

\bibitem{fano1961effects}
Ugo Fano.
\newblock Effects of configuration interaction on intensities and phase shifts.
\newblock {\em Physical review}, 124(6):1866, 1961.

\bibitem{lv2016analysis}
Bo~Lv, Rujiang Li, Jiahui Fu, Qun Wu, Kuang Zhang, Wan Chen, Zhefei Wang, and
  Ruyu Ma.
\newblock Analysis and modeling of fano resonances using equivalent circuit
  elements.
\newblock {\em Scientific reports}, 6(1):31884, 2016.

\bibitem{10.1007/978-3-030-16770-7_14}
M.~Ilchenko and A.~Zhivkov.
\newblock Bridge equivalent circuits for microwave filters and fano resonance.
\newblock In Mykhailo Ilchenko, Leonid Uryvsky, and Larysa Globa, editors, {\em
  Advances in Information and Communication Technologies}, pages 278--298,
  Cham, 2019. Springer International Publishing.

\bibitem{molina2022fano}
Mario~I Molina.
\newblock Fano resonances in an electrical lattice.
\newblock {\em Physics Letters A}, 428:127948, 2022.

\bibitem{cserti2011uniform}
J{\'o}zsef Cserti, G{\'a}bor Sz{\'e}chenyi, and Gyula D{\'a}vid.
\newblock Uniform tiling with electrical resistors.
\newblock {\em Journal of Physics A: Mathematical and Theoretical},
  44(21):215201, 2011.

\end{thebibliography}

\bibliographystyle{unsrt}
\end{document}